\documentclass[letterpaper,12pt]{article}
\usepackage{libertine}
\usepackage{graphicx}
\usepackage{amsmath}
\usepackage{amsthm}
\usepackage{amsfonts}
\usepackage{datetime}
\usepackage[linesnumbered,ruled,vlined]{algorithm2e}
\usepackage[margin=0.8in]{geometry} % Set margins to 1 inch
\usepackage{lipsum} % For generating sample text
\usepackage{graphicx} % For including images
\usepackage{hyperref} % For hyperlinks
\usepackage{float}

\newtheorem{theorem}{Theorem}[section]

\newdateformat{monthyeardate}{%
  \monthname[\THEMONTH], \THEYEAR}

\title{Eden: A Provably Secure, Ultra-Fast, and Fully Decentralized Blockchain Interoperability Protocol}

\author{
    Ke Liang \\ 
    SparkleX Team
}

\date{\monthyeardate\today} % Use today's date

\begin{document}

\maketitle

\begin{abstract}
\noindent
As the blockchain ecosystem grows and diversifies, seamless interoperability between blockchain networks has become essential. Interoperability not only enhances the usability and reach of individual chains but also fosters collaboration, unlocking new opportunities for decentralized applications. In this paper, we introduce Eden, the parallel-verified messaging protocol powering SparkleX. Eden is an elastic, decentralized envoy network built on a zero-knowledge MapReduce framework (\textit{i.e.}, ZK-MapReduce), enabling ultra-fast, secure, and fully decentralized cross-chain communication. We explore Eden’s design, its robust security model, and the innovative mechanisms that ensure its elasticity and resilience, even in demanding network environments.
\end{abstract}

\section{Introduction}
The introduction of blockchain technology, starting with Bitcoin~\cite{nakamoto2008bitcoin} in 2009, revolutionized digital transactions by enabling a peer-to-peer electronic cash system. Built on distributed ledger technology, Bitcoin ensured transparency and immutability but focused solely on decentralized digital currency. Ethereum~\cite{wood2014ethereum} later expanded blockchain’s capabilities by introducing smart contracts, unlocking decentralized applications (dApps) across industries. However, Ethereum’s scalability issues, such as slow transactions and high fees, remain a bottleneck for broader adoption.

Layer-2 solutions, including Arbitrum~\cite{kalodner2018arbitrum}, address these challenges by enhancing transaction speeds and reducing costs through strategies like rollups and sidechains~\cite{gangwal2023survey}. Yet, as these solutions proliferate, seamless interoperability~\cite{belchior2021survey} becomes critical for integrating diverse networks into a unified Web3 ecosystem. Interoperability allows data and value exchange across chains, but current cross-chain solutions face issues such as security vulnerabilities, limited scalability, and centralized dependencies.

To tackle these challenges, we introduce Eden, an innovative and elastic decentralized network of envoys designed to redefine cross-chain communication. Powered by a zero-knowledge MapReduce framework, Eden enables ultra-fast, secure, and trustless interactions across blockchains. Its parallel and non-interactive architecture ensures seamless scalability and efficiency, addressing the limitations of current solutions while setting a new benchmark for interoperability in the blockchain ecosystem.

Eden is essentially a PoS-based interoperability protocol that removes the need for traditional consensus to verify messages. Instead, validators—called envoys—operate privately and independently, voting on committed messages from the source chain. Like “mappers” in a MapReduce framework~\cite{dean2008mapreduce}, envoys verify messages in parallel and send their votes to SparkleX’s X-Chain\footnote{SparkleX’s X-Chain is an L2 EVM-based blockchain that utilizes the SparkleX Token as its gas token.}, where reducer contracts are hosted. Once enough votes are received, the message is confirmed. Thanks to this innovative design, Eden delivers superior performance compared to existing solutions in the following key areas:

\begin{itemize}
  \item \textbf{Security}. 
  As a PoS-based protocol, Eden ensures cross-chain message integrity by requiring an attacker to control over 1/3 of the total stakes to compromise the system. What sets Eden apart is its unique approach: envoys (validators) verify messages privately and independently through a probabilistic weighted mechanism. While vote weights are proportional to stakes, they are randomized using ZK-VRF~\cite{micali1999verifiable}. This innovative design enhances resilience against attacks and system disruptions, maintaining stability even if high-stake envoys fail unexpectedly.

  \item \textbf{Efficiency}.
  Eden utilizes a novel probabilistic weighted mechanism, combining ZK-VRF and the inverse transform method~\cite{lehtinen1989linear} to efficiently calculate and prove votes for each message. This reduces both the complexity and processing time typically associated with proof generation and vote calculation. Additionally, Eden employs error-correcting codes~\cite{shokrollahi2011raptor}, enabling envoys to segment messages into slices and transmit them proportionally to their weighted votes. This approach minimizes transmission latency by reducing duplicate message propagation while maintaining a fully decentralized and efficient network.

  \item \textbf{Decentralization}. 
  Eden relies on an open network of envoys—decentralized validators who independently validate and vote on cross-chain messages. Unlike traditional PoS-based interoperability protocols, envoys operate without communication or consensus requirements, ensuring true decentralization and optimal efficiency. 

\end{itemize}

The structure of this paper is as follows: Section~\ref{sec:relatedwork} reviews existing interoperability protocols, building on the foundational assumptions outlined in Section~\ref{sec:assumption}. Section~\ref{sec:overview} provides an in-depth overview of Eden, detailing its operational principles and functionalities. Section~\ref{sec:zkmapreduce} explores the ZK-MapReduce framework, focusing on its security mechanisms and guarantees. Finally, the paper concludes with key insights and findings in the closing section.

\section{Related Work}\label{sec:relatedwork}
This paper excludes centralized cross-chain solutions from its scope, as they fundamentally conflict with the core principles of blockchain technology—security, openness, and decentralization. These solutions rely on trust in a central authority, undermining the very ethos of a decentralized ecosystem.

Existing cross-chain technologies can be broadly classified into three categories: 1) Committee-Based Schemes, 2) Light-Client Schemes, and 3) Multi-Signature Schemes.
\begin{itemize}
  \item \textbf{Committee-Based Scheme}. Committee-based cross-chain solutions authenticate transactions through a group of committees using Byzantine Fault-Tolerant (BFT) consensus algorithms~\cite{zhang2022reaching}. Examples include Axelar~\cite{Axelar2021}, PolyNetwork~\cite{li2022polybridge}, LayerZero~\cite{zarick2021layerzero}, and CCIP~\cite{Chainlink2021}. While these systems are relatively simple to implement and avoid complex proof-of-work computations, they suffer from significant communication overhead, limited scalability across multiple chains, and a fundamental misalignment with the blockchain’s core principle of decentralization.

  \item \textbf{Light-Client Scheme}. Light-client schemes, like those in~\cite{westerkamp2020zkrelay} and~\cite{xie2022zkbridge}, use zero-knowledge proofs (e.g., zkSNARKs) to validate cross-chain messages. Off-chain servers generate proofs of state changes on the source blockchain, which are verified by smart contracts on the destination chain. However, proof generation is computationally expensive, taking over 20 seconds on high-end machines~\cite{xie2022zkbridge}, and verification incurs high gas costs. These challenges limit the practicality of this approach, especially as the number of interconnected blockchains grows.

  \item \textbf{Multi-Signature Scheme}. The simplest approach is Multi-Signature schemes, like Hyperlane, where a group of validators approves messages for the destination chain. Once enough signatures are collected, the message is committed. However, relying on centralized signature storage and message delivery makes this approach less secure and prone to attacks.
\end{itemize}

While these approaches have advanced cross-chain communication, they all face challenges like centralization, inefficiency, and limited scalability. These shortcomings underscore the need for a more robust, decentralized, and efficient solution tailored to the demands of a multi-chain future. Eden fills this gap with its innovative design, setting a new standard for cross-chain interoperability.

\section{Assumption} \label{sec:assumption}
Eden is built on standard cryptographic principles, utilizing public-key signatures and hash functions to ensure security and reliability. The protocol assumes that the majority of envoys—participants who validate and vote on messages—are honest and operate error-free software. A key element of Eden’s security is the assumption that the total stake held by these honest envoys, denominated in SparkleX tokens, exceeds a critical threshold $h$, which is set above two-thirds of the total stake. This threshold ensures that any adversary would need to control a substantial portion of the network’s financial resources to launch a successful attack, making such scenarios highly improbable.

Eden also accounts for the potential risk of individual envoys being compromised. However, the system is designed with the assumption that it is practically impossible for an adversary to corrupt enough envoys to bring the total honest stake below the critical threshold $\tau$. This robust design provides resilience against concentrated attacks, ensuring that the network remains secure and maintains its integrity under adverse conditions.

\section{Overview} \label{sec:overview}
\begin{figure}[H]
  \centering
  \includegraphics[width=0.6\textwidth]{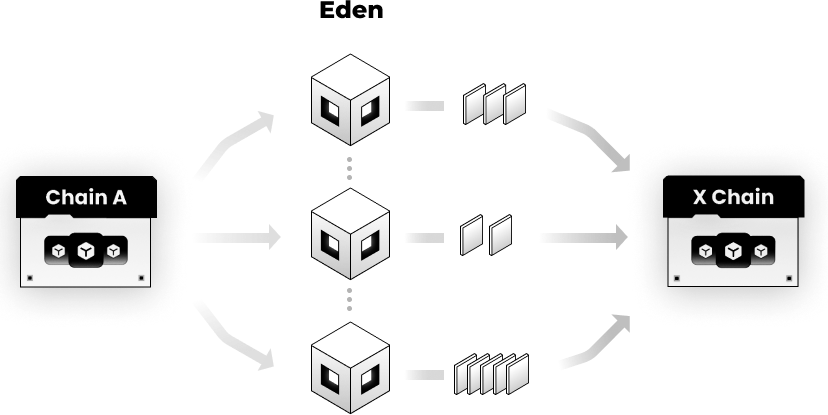}
  \caption{Eden consists of a cluster of envoys, each independently and non-interactively delivering cross-chain messages.}
  \label{fig:eden-overview}
\end{figure}

As shown in Figure~\ref{fig:eden-overview}, the Eden comprises specialized nodes called \textbf{envoys}. These envoys are responsible for retrieving the latest committed updates—such as the liquidity information of SparkleX users—from multiple blockchains. By leveraging light clients or relayers, envoys ensure this process is both efficient and reliable. This design allows Eden to enable SparkleX’s X-Chain to consistently maintain accurate and up-to-date liquidity information for all users across connected networks.

We introduce an innovative ZK-MapReduce framework for Eden to authenticate messages between the source blockchain and the target blockchain (e.g., SparkleX’s X-Chain). Eden utilizes a \textit{cryptographic sortition mechanism}, where each envoy independently generates a weighted vote for the message, with the weight determined by the envoy’s stake on the X-Chain. This design ensures the security guarantees of PoS-based protocols in a fully \textit{non-interactive} manner, eliminating the need for communication between envoys. As a result, cross-chain messages are verified and delivered in parallel, reducing latency and maximizing scalability as the number of source chains increases.

To reduce transmission latency between envoys and the destination chain, envoys send only a portion of the cross-chain message along with the necessary proofs. To enable efficient data recovery from these partial transmissions, Eden employs Raptor Codes, an advanced form of Fountain Codes. The message is encoded into $\tau$ distinct pieces (where $\tau$ is a predefined value discussed in Section~\ref{section:rnd}), and a subset of v pieces (corresponding to the number of votes) is randomly selected and transmitted to the destination chain. This approach ensures both efficient data transmission and reliable data recovery on the destination chain, maintaining robustness and scalability.

On the destination chain (i.e., X-Chain), a designated reducer handles the verification and decoding of messages submitted by the envoys. This process begins once the reducer receives and assembles $\tau$ message segments, each accompanied by its respective zk-proof. By validating the proofs and reconstructing the original message, the reducer ensures a secure, efficient, and reliable method for processing cross-chain communications. This design guarantees the integrity and authenticity of the transmitted data while maintaining scalability and robustness.

In summary, Eden’s innovative architecture combines the ZK-MapReduce framework, probabilistic weighted voting, and advanced error-correcting techniques to deliver a highly efficient and secure cross-chain messaging protocol. By leveraging decentralized envoys, non-interactive validation, and scalable data transmission, Eden ensures seamless interoperability between blockchains. This robust design not only minimizes latency and maximizes scalability but also aligns with the principles of decentralization, making it a foundational element of SparkleX’s omnichain liquidity network.

\section{ZK-MapReduce Framework}\label{sec:zkmapreduce}
\begin{figure}[H]
  \centering
  \includegraphics[width=1\textwidth]{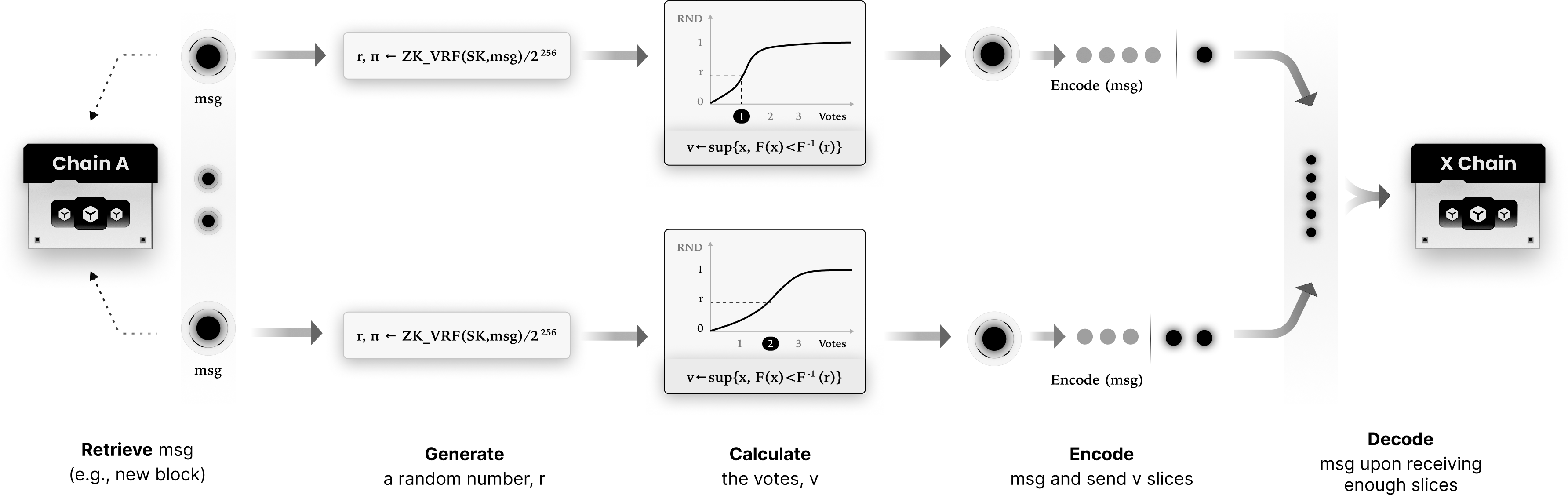}
  \caption{ZK-MapReduce framework.}
  \label{fig:architecture}
\end{figure}

Eden’s ZK-MapReduce framework, as shown in Figure~\ref{fig:architecture}, is at the core of its innovative interoperability protocol. This framework is designed to ensure efficient, secure, and decentralized message verification and transmission between blockchains. By combining the principles of the MapReduce computational model with zero-knowledge cryptography, ZK-MapReduce allows envoys to independently and non-interactively verify cross-chain messages while maintaining the integrity and authenticity of the data. This section provides an overview of the ZK-MapReduce framework, explaining its core components and functionalities. We discuss the mapping and reducing processes, the use of zero-knowledge proofs, and the mechanisms that enable scalability and robustness.
\subsection{Mapping Phase}
% In the mapping phase, envoys independently verify and vote on cross-chain messages in a fully non-interactive manner. Using a combination of a cryptographically secure random number and the inverse transform method, each envoy calculates a vote weight proportional to their stake on the X-Chain. This probabilistic weighted mechanism ensures fairness and aligns with the principles of PoS-based protocols.
The mapping phase is where envoys independently verify and vote on cross-chain messages in a fully non-interactive and decentralized manner. By combining cryptographically secure random numbers with the inverse transform method, each envoy calculates a vote weight that is proportional to their stake on the X-Chain. This probabilistic mechanism ensures fairness while maintaining the security guarantees of PoS-based protocols.

Here’s how the process works: 1.) \textbf{Message Retrieval:} Envoys first retrieve the latest committed updates from the source blockchain, such as cross-chain messages or liquidity data, ensuring they have the most accurate information. 2). \textbf{Vote Calculation}: Each envoy generates a secure random number using zk-VRF and applies the inverse transform method to calculate its vote weight. This weight is proportional to the envoy’s stake but incorporates randomness, preventing deterministic outcomes and enhancing resilience against potential adversaries. 3). \textbf{Encoding and Transmission}: Instead of transmitting the full cross-chain message, envoys encode the message into $\tau$ segments using Raptor Codes. Based on their calculated vote weight, envoys send a corresponding number of these segments, along with their zk-proof, to the destination chain.

The message retrieval and the encoding and transmission steps are relatively straightforward processes. In this section, we will concentrate on the more intricate and critical step of vote calculation within the mapping phase.

% Building on the initial overview, this section delves deeper into the intricacies of Eden's design, offering a comprehensive exploration of its unique mechanisms and architectural choices. We will also provide a thorough analysis of the security guarantees inherent in Eden's structure, affirming its robustness and reliability in the blockchain ecosystem.

\subsection{Vote Calculation}\label{section:voting}
Vote calculation is a crucial component of the mapping phase, combining randomness and proportionality to ensure both fairness and security. Eden employs a decentralized approach where envoys are selected to verify cross-chain messages based on their stakes, effectively mitigating Sybil attacks. The probability of selection and voting power of each envoy are directly proportional to their staked tokens, ensuring the network’s integrity and resilience against adversaries with minimal stakes while maintaining a fair and balanced validation process.

Let $E = \bigcup_{i \in \mathbb{Z}^+} \{e_i\}$ denote the set of all envoys, and $S = \bigcup_{i \in \mathbb{Z}^+} \{s_i\}$ represent the set of their stake holdings, respectively. Each envoy, denoted as $e_i$, performs a Bernoulli trial for each unit of their stake, with a probability of success $p$. The probability of envoy $e_h$ being selected is given by $1 - (1 - p)^{s_i}$, and their expected weight of vote is $p \times s_i$. Thus, the weight of vote for an envoy $e_i$, denoted as $v_i$, is a random variable that follows a Binomial distribution, i.e., $v_i \sim B(s_i, p)$. The cumulative distribution function (CDF) for $v_i$ is defined as:
\begin{align}
  F(k; s_i, p) &= \Pr(v_i \leq k) \\
  &= \sum_{j=0}^{k} \binom{s_i}{j} p^j (1 - p)^{s_i - j}
\end{align}

\begin{figure}[H]
  \centering
  \includegraphics[width=0.5\textwidth]{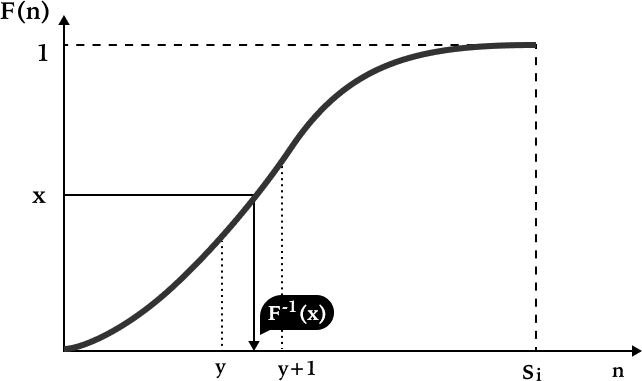}
  \caption{Determining the votes of $e_i$ via Inverse Transform Method: Here, $x$ represents a uniformly distributed random number in the interval $[0,1]$. $y$ is defined as the greatest integer less than or equal to $F^{-1}(x)$. Consequently, the voting power of $e_i$ is denoted by $y$.}
  \label{fig:vote-calculation}
\end{figure}

To calculate the voting weight \( v_i \) for each envoy \( e_i \) in the set \( E \), given their stake \( s_i \) and the probability \( p \), we employ a straightforward algorithm based on the inverse transform method. This is outlined in Figure~\ref{fig:vote-calculation} and Algorithm~\ref{algo:voting}. 

The procedure begins with each envoy generating a random number, denoted as \( x \) (detailed in Section~\ref{section:rnd}), uniformly distributed over the interval \( (0, 1) \) (refer to Line 1 in Algorithm 1). The algorithm then compares \( x \) with \( F(0; s_i, p) = (1 - p)^{s_i} \) — the probability that none of the \( s_i \) tokens is selected. If \( x \) is not greater, the process halts and returns 0, as detailed in Lines 2-3 of Algorithm~\ref{algo:voting}. 

Otherwise, the algorithm computes the voting weight \( v_i \) using the inverse cumulative distribution function (CDF), \( F^{-1}(x) \). This computation leverages the random number \( x \), the envoy’s stake \( s_i \), and the probability \( p \), effectively determining \( v_i \) in a probabilistic and efficient manner. This method ensures a fair and proportional allocation of voting weights while maintaining computational simplicity.

% The inverse Cumulative Distribution Function (CDF), \( F^{-1}(x) \), is subsequently used to compute the voting power \( v_i \), based on the generated random number \( x \), and the specified \( s_i \) and \( p \).

\begin{algorithm}
  \caption{Calculate \( v_i \) for envoy \( e_i \), given \( p \)}\label{algo:voting}
  Generate a random number \( x \sim U(0, 1) \);\\
  \If{\( x \leq (1 - p)^{s_i} \)}{
      Return 0;\\
  }
  Return \( \sup\{v_i \in \mathbb{Z}^+ : F(v_i) \leq F^{-1}(x)\} \);\\
\end{algorithm}

\subsection{Random Number Generation}\label{section:rnd}
Every envoy uses Verifiable Random Functions (VRFs) to generate random numbers for calculating their voting weight for each cross-chain message, as illustrated in Line 1 of Algorithm~\ref{algo:voting}. The use of VRFs ensures efficient and secure verification of these random numbers by the reducer. Furthermore, VRFs enable envoys to compute their voting weight independently and non-interactively, eliminating the need for coordination with other envoys.

% Every envoy employs Verifiable Random Functions (VRFs) to generate random numbers for calculating their voting power for every cross-chain message, as shown in Line 1 of Algorithm~\ref{algo:voting}. The use of VRFs facilitates efficient verification of these numbers by the reducer. Additionally, VRFs enable envoy to independently and non-interactively compute their voting power. 

Let $\mathrm{SK}_i$ be the private key of envoy $e_i$\, and let $\mathrm{HASH}(msg)$ be the hash of the cross-chain message $msg$. Using VRFs, every envoy $e_i$ generates a tuple as follows:
\begin{align}
  (x_i, \pi_i) = \mathrm{VRF}(\mathrm{SK}_i, \mathrm{HASH}(msg))/2^{256}
\end{align}

where $x_i$ is the random number uniformly distributed over  $(0, 1)$, used to calculate the voting weight. $\pi_i$ is the proof associated with the random number $x_i$, ensuring that it was correctly generated using $\mathrm{SK}_i$ and $\mathrm{HASH}(msg)$. This tuple $(x_i, \pi_i)$ allows the reducer to verify the correctness of the random number $x_i$ using the corresponding public key of the envoy and the message hash. This design ensures the integrity and fairness of the voting process, making it secure and tamper-proof.

\subsection{Security Guarantee}\label{section:security}
The security of Eden’s protocol is designed to align with the established guarantees of PoS protocols, ensuring that cross-chain communication remains robust, decentralized, and resistant to adversarial attacks. This subsection demonstrates how Eden achieves equivalent security to PoS by leveraging a probabilistic voting mechanism, zk-VRF, and stake-based selection. The following analysis outlines the key components and principles that uphold Eden’s security model.

Given the impracticality of accurately determining the exact number of active envoys  \( |E| \) at any given time, Eden adopts a predetermined value for $p$ that remains unaffected by dynamic and unpredictable factors. Specifically, $p$ is defined as \( \tau/K \), where \( K \) denotes the total token supply\footnote{SparkleX token has a fixed maximum supply.}, and \( \tau \) represents the number of votes required by the reducer on the destination chain to confirm each cross-chain message. The choice of \( \tau \) is a critical parameter, as it directly influences both the efficiency and security of Eden’s protocol. A higher \( \tau \) increases security by requiring more votes for confirmation but can impact efficiency by raising the computational and communication overhead. The implications and considerations for selecting \( \tau \) will be further analyzed in this section.

It is important to note that within this paper, the terms \textit{selected tokens} and \textit{weights of votes} are used interchangeably. This usage stems from the implementation of a zk-VRF based sortition mechanism, which effectively simulates the random selection of $\tau$ tokens from a pool of $K$ tokens, achieving this in a fully decentralized and non-interactive manner.

Let \( S_h \) and \( S_a \) represent the total number of tokens staked by online honest envoys and adversaries, respectively. The honesty threshold, denoted as \( h \), corresponds to the fraction of tokens staked by honest envoys, such that \( S_h = h(S_h + S_a) \). Additionally, let \( \alpha \)  represent the \textbf{activity} level of the system, defined as the fraction of tokens held by online envoys. Thus, the total number of online tokens can be expressed as \( (S_h + S_a) = \alpha K \), where  \(K\) is the total stake in the system.

Define \( X_h \) and \( X_a \) as the number of tokens selected from honest envoys and adversaries, respectively, based on Algorithm 1. As described in Section~\ref{section:voting}, \( X_h \) and \( X_a \) follow Binomial distributions: \( X_h \sim B(S_h, p) \) and \( X_a \sim B(S_a, p) \), where \(p\) is the probability of a token being selected in a Bernoulli trial.

Given the assumption that \( S_h \) and \( S_a \) are sufficiently large\footnote{This implies $S_hp>5$, and $S_ap>5$.}, the Binomial distributions \( B(S_h, p) \) and \( B(S_a, p) \) can be closely approximated by Normal distributions as follows:

% Let \( S_h \) and \( S_a \) denote the total number of tokens owned by the online honest envoys and adversaries, respectively. The honesty threshold, indicated by \( h \), represents the fraction of tokens that honest envoys stake, such that \( S_h = h(S_h + S_a) \). Let \( \alpha \) be the \textbf{activity} \footnote{Active envoys can send votes or messages to the receiver successfully.} level of the system, defined as the fraction of tokens held by online envoys. Hence, we have \( (S_h + S_a) = \alpha K \).

% Define \( X_h \) and \( X_a \) as the number of tokens selected from honest nodes and adversaries, respectively, as determined through Algorithm 1. As discussed in Section~\ref{section:voting}, \( X_h \) and \( X_a \) are modeled by Binomial distributions, with \( X_h \sim B(S_h, p) \) and \( X_a \sim B(S_a, p) \). Given the assumption that \( S_h \) and \( S_a \) are sufficiently large\footnote{Which means $S_hp>5$, and $S_ap>5$}, both \( B(S_h, p) \) and \( B(S_a, p) \) can be closely approximated with Normal distribution as follows:
\begin{align}
  X_h &\sim \mathcal{N}(\mu_h, \sigma_h^2),  \mu_h = S_h p, \sigma_h^2 = S_h p (1 - p) \label{eq:honest-dist} \\
  X_a &\sim \mathcal{N}(\mu_a, \sigma_a^2),  \mu_a = S_a p, \sigma_a^2 = S_a p (1 - p) \label{eq:adversarial-dist}
\end{align}

To enable efficient verification of cross-chain messages by the receiver on the destination chain without requiring the full receipt of all message slices, a predefined threshold fraction of voting weights, denoted as \( \theta \), is established. This threshold ensures that a cross-chain message is considered valid with a very high probability (\( > 1 - 10^{-10} \)) if it gathers more than \( \theta \tau \) votes from envoys. This approach balances efficiency and reliability, reducing overhead while maintaining strong security guarantees.

% As a result, \( \Pr(X_h < \theta \tau) \) should be negligible. Additionally, it is equally crucial to maintain that the probability \( \Pr(X_a \geq \theta \tau) \) is minimal. This condition ensures a significantly lower risk of fraudulent activities by adversaries, even in scenarios characterized by a relatively low level of system activity, denoted as \( \alpha \). 

\begin{theorem}\label{theorem:honest}
Let \( h \) represent the fraction of tokens held by honest envoys, and \( \alpha \) denote the proportion of tokens owned by active envoys. Suppose the probability of a token being selected is defined as \( p = \tau / K \), where \( \tau \) is a predefined constant and \( K \) is the maximum token supply. The cross-chain message will be confirmed at the destination chain \textit{with high probability (w.h.p.)} upon receiving \( \theta \tau \) votes, provided the following condition is satisfied:

  \begin{align}
    \theta < h\alpha - 6.36 \sqrt{ \frac{h\alpha}{ \tau} } \label{inq:theta}
    \end{align}
\end{theorem}

  \begin{proof}
    Referring to Equation~\ref{eq:honest-dist}, we note that \( X_h \sim B(S_h, p) \), where \( S_h \) denotes the aggregate number of tokens owned by active honest envoys. In order to guarantee that the receiver on the destination chain accrues at least \( \theta\tau \) votes for the cross-chain message, the ensuing criterion must be met:

    \begin{align}
      \Pr(X_h < \theta\tau) &= \Pr\left(Z < \frac{\theta\tau - \mu_h}{\sigma_h}\right) < 10^{-10}, \label{inq:honest-dist}
    \end{align}
    where the random variable $Z$ follows a standard normal distribution, denoted as $Z \sim \mathcal{N}(0, 1)$. It is important to note the extremely low probabilities at the tails of this distribution: both $\Pr(Z > 6.36)$ and $\Pr(Z < -6.36)$ are less than $10^{-10}$. These probabilities reflect the rarity of extreme deviations from the mean in a standard normal distribution and are pivotal in the ensuing derivations and conclusions. Thus the following shall hold to satisfiy Inequation~\ref{inq:honest-dist} 
    \begin{align}
      &&\frac{\theta\tau - \mu_h}{\sigma_h} &< -6.36 \\
      &\Rightarrow& \theta &< h\alpha - 6.36 \sqrt{ \frac{h\alpha}{ \tau} (1-p)}
    \end{align}
    
    In practice, $p < 0.001$, hence we have $\theta < h\alpha - 6.36 \sqrt{ \frac{h\alpha}{ \tau} }$. 
  \end{proof}

  % We now establish that the robust security guarantee of the system is maintained as long as the share of tokens controlled by malicious envoys does not exceed the threshold $h$, irrespective of the number of such malicious envoys.

  We now demonstrate that the system’s robust security guarantees remain intact as long as the proportion of tokens controlled by malicious envoys does not exceed the critical threshold $ h $, regardless of the number of malicious envoys. This result highlights the protocol’s resilience to adversarial behavior, as it ensures that the integrity of the system is determined by the distribution of stakes rather than the sheer number of malicious participants.

  \begin{theorem}\label{theorem:adversarial}
    Let \( h, \alpha, \tau, p, K \) be as defined in Theorem~\ref{theorem:honest}. We assert that no fraudulent cross-chain message will be confirmed at the destination chain \textit{with high probability (w.h.p)} under the condition that the following holds:

    \begin{align}
      \theta > (1 - h)\alpha + 6.36\sqrt{\frac{(1 - h)\alpha}{\tau}} \label{inq:tau2} 
      \end{align}
  \end{theorem}
    
    \begin{proof}
      Referring to Equation~\ref{eq:adversarial-dist}, we observe that \( X_a \sim B(S_a, p) \), where \( S_a \) is the total number of tokens held by active malicious envoys. To ensure that the receiver on the destination chain does not accumulate \( \theta\tau \) votes for the cross-chain message, the following condition must be satisfied:

      \begin{align}
        \Pr(X_a \geq \theta\tau) &= \Pr\left(Z \geq \frac{\theta\tau - \mu_a}{\sigma_a}\right) < 10^{-10} \label{inq:adversarial-dist}
      \end{align}
      where the random variable $Z$ follows a standard normal distribution, denoted as $Z \sim \mathcal{N}(0, 1)$. Considering  $\Pr(Z > 6.36) < 10^{-10} $, we have the following: 
      \begin{align}
        &&\frac{\theta\tau - \mu_a}{\sigma_a} &> 6.36 \\
        &\Rightarrow& \theta &>(1 - h)\alpha + 6.36\sqrt{\frac{(1 - h)\alpha}{\tau} (1-p)}  
      \end{align}
      In practice, $p < 0.001$, hence we have $ \theta > (1 - h)\alpha + 6.36\sqrt{\frac{(1 - h)\alpha}{\tau}}$, and this prove the Theorem~\ref{theorem:adversarial}  
    \end{proof}

% \begin{align}
%   \Pr(X_a \geq \theta\tau) &= \Pr\left(Z \geq \frac{\theta\tau - \mu_a}{\sigma_a}\right) < 10^{-10} \\
%   \Pr(X_h < \theta\tau) &= \Pr\left(Z < \frac{\theta\tau - \mu_h}{\sigma_h}\right) < 10^{-10},
% \end{align}
% whre the random variable $Z$ follows a standard normal distribution, denoted as $Z \sim \mathcal{N}(0, 1)$. It is important to note the extremely low probabilities at the tails of this distribution: both $\Pr(Z > 6.36)$ and $\Pr(Z < -6.36)$ are less than $10^{-10}$. These probabilities reflect the rarity of extreme deviations from the mean in a standard normal distribution and are pivotal in the ensuing derivations and conclusions. Thus we have the following inequations: 
% \begin{align}
% (1 - h)\alpha + 6.36\sqrt{\frac{(1 - h)\alpha}{\tau}}  < \theta < h\alpha - 6.36 \sqrt{ \frac{h\alpha}{ \tau}}
% \end{align}
 
\subsection{Choosing Optimal Parameters}
In Eden, selecting the parameters $ \theta $ (the fraction of voting weights required for message validation) and $ \tau $ (the total number of votes required for the reducer to confirm a message) is crucial for balancing security and efficiency. This section derives the appropriate values of $ \theta $ and $ \tau $ based on the protocol’s security assumptions and activity levels, ensuring the system’s robustness without excessive communication overhead.

\subsubsection{Security Foundation}
Eden’s security guarantees are built on the supermajority assumption common in PoS blockchains, requiring that more than $ 2/3 $ of tokens are held by honest envoys. To formalize this, define $ Y = X_h - 2X_a $, where $ X_h $ and $ X_a $ represent the number of tokens selected from honest envoys and adversaries, respectively. $ Y $ follows a Normal distribution, $ Y \sim \mathcal{N}(\mu_y, \sigma_y^2) $, with the following parameters:

\begin{align}
\mu_y = (3h - 2)\alpha\tau, \quad \sigma_y^2 = (4 - 3h)\alpha\tau,
\end{align}

where $ h $ is the fraction of tokens held by honest envoys, and $ \alpha $ is the fraction of tokens held by active envoys. To maintain security, the probability $ \Pr(Y \leq 0) $ must remain negligible, ensuring that honest votes dominate.

\subsubsection{Determining $\tau$}
By converting $ Y $ into a standard Normal variable $ Z \sim \mathcal{N}(0, 1) $, the probability $ \Pr(Y \leq 0) $ is expressed as:

\begin{align}
\Pr(Y \leq 0) = \Pr \left( Z \leq -\frac{\mu_y}{\sigma_y} \right) = \Pr \left( Z \leq -\frac{(3h - 2)\alpha\tau}{\sqrt{(4 - 3h)\alpha\tau}} \right).
\end{align}

Considering \(\Pr(Z \leq -6.36) \leq 10^{-10}\), we have

\begin{align}
\tau > \frac{40.5(4 - 3h)}{(3h - 2)^2 \alpha}. \label{inq:tau}
\end{align}

Bitcoin, one of the most secure and widely adopted blockchains, is known to be vulnerable to certain attacks, such as selfish mining, when adversaries control more than 25\% of the network’s hashrate. To ensure Eden is resilient to similar threats, we postulate that the honest threshold $ h $ must exceed 0.75. This means at least 75\% of the tokens in the system must be held by honest envoys. Given that the right-hand side of Inequation~\ref{inq:tau} is a monotonic, non-increasing function of $ h $, this implies that:

\begin{align}
\tau > \frac{1134}{\alpha}.
\end{align}

This inequality ensures that $ \tau $ is sufficiently large to accommodate varying activity levels ( $\alpha$ ) while maintaining robust security.

\subsubsection{Determining $\theta$}
The parameter $ \theta $ must ensure that the honest votes dominate the total received votes, even under adversarial conditions. Based on Eq.~\ref{inq:tau2}, the lower bound of $ \theta $ is given:

\begin{align}
\theta > (1 - h)\alpha + 6.36\sqrt{\frac{(1 - h)\alpha}{\tau}}.
\end{align}

For $ h \geq 0.75 $ and $ \alpha \in [0, 1] $, this simplifies further to:

\begin{align}
\theta > \frac{1}{4} + \frac{3.18}{\sqrt{\tau}}.
\end{align}

This inequality establishes a threshold for $ \theta $, ensuring that the system remains secure against adversaries while minimizing the number of votes required.

\subsubsection{Parameter Selection}
In Eden, we set $ \tau = 5000 $ and $ \theta = 0.3 $, balancing efficiency and security. Substituting these values into Inequality~\ref{inq:theta} yields:

\begin{align}
h\alpha - 6.36\sqrt{\frac{h\alpha}{5000}} > 0.3.
\end{align}

This condition is satisfied if $ \alpha > 0.53 $ and $ h \geq 0.75 $. Under these parameters:
\begin{itemize}
    \item A cross-chain message is confirmed with high probability if it receives at least 1500 votes ( $\theta \tau = 0.3 \times 5000 $).
    \item The system ensures resilience against adversaries, as long as more than 75\% of tokens are held by honest envoys and the activity level exceeds 53\%.
\end{itemize}
By setting $ \tau = 5000 $ and $ \theta = 0.3 $, Eden achieves a secure and efficient balance. These values guarantee high-probability confirmation of cross-chain messages while minimizing communication overhead. The protocol’s ability to adapt to varying activity levels and maintain security under supermajority assumptions highlights its robustness and scalability.

\subsection{Reducing Phase}
The reducing phase occurs on the destination blockchain (i.e., X-Chain), where a designated reducer is responsible for processing and verifying the transmitted data, such as liquidity information. Hosted on X-Chain, the reducer ensures the integrity, validity, and completeness of the received votes and message segments. This phase serves as the final step in confirming cross-chain messages, maintaining the protocol’s robust security and operational efficiency guarantees.

As outlined in Algorithm~\ref{algo:reduction}, the reducer’s primary task is to authenticate the tuples received from envoys, which consist of encoded message segments and their corresponding proofs. After successful verification, the reducer appends the validated segments for further processing. The decoding process, which reconstructs the original cross-chain message, begins once the total size of the accumulated encoded segments exceeds the predefined threshold $\theta\tau$. This ensures that only secure and complete messages are processed and committed to the destination chain. Here is how reducer works:
\begin{itemize}
    \item \textbf{Vote Verification}: The reducer collects the votes $ v_i $ and accompanying zk-proofs $ \pi_i $ submitted by the envoys. Each proof validates the envoy’s eligibility and the correctness of its vote calculation, ensuring compliance with the protocol’s rules. This process includes: 1). Verifying that the vote weight $ v_i $ was derived using the envoy’s stake and zk-VRF-generated random number. 2). Ensuring that the cumulative weight of valid votes meets or exceeds the threshold $ \theta\tau $ for message confirmation.
    \item \textbf{Message Decoding}: Envoys transmit encoded message segments using Raptor Codes during the Mapping Phase. The reducer assembles these segments to reconstruct the original cross-chain message. This involves: 1). Accumulating $\tau$ segments from the submitted slices. 2). Decoding the message using the Raptor Code decoding algorithm, which efficiently reconstructs the original data even if some segments are lost or delayed.
    \item \textbf{Message Commitment}: Once the original message is reconstructed and verified, the reducer commits the message to the destination chain. This commitment ensures that the cross-chain operation is finalized and accessible to all parties interacting with the destination chain.
\end{itemize}

The Reducer Phase is the final step in Eden’s ZK-MapReduce framework, ensuring that cross-chain messages are securely verified, decoded, and committed. By combining cryptographic proofs, threshold voting, and robust error correction, the reducer achieves both high security and operational efficiency. This phase is pivotal in enabling Eden to support seamless and reliable cross-chain interoperability in SparkleX’s omnichain liquidity network.

% This phase ensures the integrity and reliability of the cross-chain communication while maintaining low computational overhead on the destination chain.

% \subsection{Reduction}
% As outlined in Algorithm~\ref{algo:reduction}, the receiver's primary task is to authenticate the tuples received from the envoys, followed by appending the encoded message segments post-verification. The decoding process, aimed at reconstructing the original cross-chain message, commences once the aggregate size of these received encoded segments surpasses a predetermined threshold, specifically $\theta\tau$.

\begin{algorithm}
  $\mathrm{PK}_i \gets$ the public key of envoy $i \in E$\;
  $\mathrm{receivedSlices}[] \gets$ array of slices received\;
  
  \While{receive \{$\mathrm{encodedMsg}[:v_i], v_i, \pi_i$\} from envoy $i\in E$}{
  \If{$\mathrm{Verify}(v_i, \mathrm{PK}_i, \pi_i)$ is false}{
  ignore and break\;
  }
  receivedSlices[].append(encodedMsg[:$v_i$]) \;
  
  \eIf{$\mathrm{receivedMsg.size}() \geq \theta\tau$}{
      $msg \gets \mathrm{raptorq.Decode}(\mathrm{receivedSlices[]}, \tau) $\;
      
  }{
      continue\;
  }
  }
  \caption{Reduction} \label{algo:reduction}
  \end{algorithm}

\section{Conclusion}
Eden is a significant breakthrough in cross-chain communication, offering a secure, decentralized, and efficient solution through its innovative ZK-MapReduce framework. Addressing the challenges of interoperability in a multi-chain ecosystem, Eden enables seamless and trustless message verification between blockchains, serving as the backbone of SparkleX’s omnichain liquidity network.

Eden’s decentralized, non-interactive design, powered by zk-VRF, eliminates the need for envoy coordination, reducing overhead and enhancing scalability. Its probabilistic weighted mechanism ensures fairness and prevents Sybil attacks by assigning vote weights proportional to envoys’ stakes while incorporating randomness for added security. The mapping and reducing phases work in harmony to achieve efficient vote generation, message encoding, and decoding, all while minimizing communication and computational costs.

The protocol adheres to the robust security standards of Proof-of-Stake systems, leveraging the supermajority assumption to ensure integrity and resistance to adversarial influence. By optimizing parameters such as $ \tau = 5000 $ and $ \theta = 0.3 $, Eden balances efficiency with strong security guarantees, confirming cross-chain messages reliably and efficiently.

Beyond messaging, Eden creates a foundation for a more interconnected blockchain ecosystem, enabling secure interoperability and driving the growth of decentralized applications and liquidity sharing. Its scalable and adaptive architecture ensures it can meet the evolving demands of blockchain networks, supporting SparkleX’s vision for a unified Web3 ecosystem.

\bibliographystyle{plain} % You can change the style
\bibliography{references} % References file

% \begin{thebibliography}{9}
% \bibitem{sampleRef}
% Author Name, \textit{Title of the Paper}, Journal Name, Volume (Year), Page numbers.
% \end{thebibliography}

\end{document}